\documentclass[12pt, doublespacing]{asme2ej}
\usepackage{setspace}   
\usepackage{epsfig} 
\usepackage{graphicx}
\usepackage{caption}
\usepackage{subcaption}

\title{Virtual Reality based Learning Systems}

\author{Yang Cheng
    \affiliation{
	Shanghai University of Science and Technology\\
    Email: chengyang532@sust.edu.cn
    }	
}

\begin{document}

\maketitle    

\begin{abstract}
This article is based on studies of the existing literature, focusing on the states-of-the-arts on virtual reality (VR) and its potential uses in learning. Different platforms have been used to improve the learning effects of VR that offers exciting opportunities in various fields. As more and more students want in a distance, part-time, or want to continue their education, VR has attracted considerable attention in learning, training, and traditional education. VR based learning enables operators to bring together all disciplinary resources in a common playground. The VR base multimedia platform has successfully demonstrated great potential of education and training. In this paper, we will discuss existing systems and their uses and address the technical challenges and future directions.
\end{abstract}

\section{Introduction}
Currently, there is a problem restricting the development of augmented reality browser, cross-platform operability including poor working mode hardware limitations and the importance of cognitive research is not enough. In order to improve the mobile augmented reality browser user experience, by analyzing the initial needs of users, to find the user's cognitive rules, this paper aims to establish a new augmented reality browser system, optimize the design of new systems. With software developed and optimized interface design, augmented reality browser user experience has been significantly improved.

However, the track mark and registered on the basis of the establishment of visualization and interaction on the basis of augmented reality. The system does not flag will stop working in a complex outdoor environments. One paper proposed a new approach, through the use of semantic web technology, in four aspects described in the context of the relationship between data and between. Using their new method to build context-aware framework for improving the efficiency push personalized service. However, the required application layer visualization of augmented reality is still further research in this paper.

Physical volume has been frequently used in the estimate body composition \cite{01}. Measurements traditionally require complex and expensive equipment companies, such as body scanning laser displacement plethysmograph \cite{02}\cite{03} underwater immersion gold \cite{04}. the use of two-dimensional pictures can help in simplifying the Measurement of the BV. photogrammetry is the use of photographic methods that can be employed to find the volume of irregularly shaped body in \cite{05}. There are two basic methods of photogrammetry of, mono- and stereo. Single refers to the use of only one camera, and stereo refers to the use of two or more cameras that can be used to capture the shoulder of depth perception. the use of such photographic method of measuring the exchange. There are ways to estimate BV  presented by the body can be represented suppose the levels ellipse and the areas and the perimeters of the ellipse can be calculated from the axis carefully measured our photo \cite{06}. Although the cross-section in a different body completely elliptical shares are not shown in later studies like yours \cite{07}, we have the oval approximation them in our study and develop an automated image processing method of the extract cross sectional oval for the hands and feet and hands trunk of the body. 

We will these pixels on the elliptical cross section of the entire volume of the body is represented in pixels. We also show that the volume expressed in this manner can be effectively used to predict body composition. Compared to other full-body scanners and other companies calculation method  \cite{08}, our method is easy to use, we only use a camera to capture images of the back and sides. It also saves complex and expensive installation cost of the system. In addition to BV, the shape and contours of the body can provide useful information on body composition. Individuals can also visually estimated using visual cues from the body  shape \cite{09}. Cornell et al. Presents a body shape assessment scale analysis of women's figures \cite{10}. Development In this study, nine body shape assessment scales, from the front and side views.

\section{The Proposed System}
Because of current hardware limitations of mobile phones, such as image recognition, image recognition and real-time needs of high-speed computing, computer side must be addressed. Mobile device over the wireless network and enhance the exchange of data between servers reality. To make up for mobile devices with limited data processing capacity, distributed architecture for mobile augmented reality system. Perform different computing tasks at the client and server respectively. Backend cloud scene offline and online learning recognition process server computing. In addition to tracking, user location, and comments have also been rendered mobile phones. Through mutual exchange and processing of data, we can provide a mixed interactive experience enables mobile phone platforms to customers. First, the real-world image sequences and camera feature points extracted from the captured image. The feature point data and the user's location is sent to the cloud server over the wireless network. Target information is to quickly identify the server in the cloud, and sent to the mobile phone sample training after completion. Finally, the real object rendering around virtual items to enhance the display. With the help of the posture sensor data, the mobile client allows accurate tracking, which means that the label is always moving virtual and the real object. Figure 1 shows the system framework balls.

\begin{figure}
    \includegraphics[width=18cm]{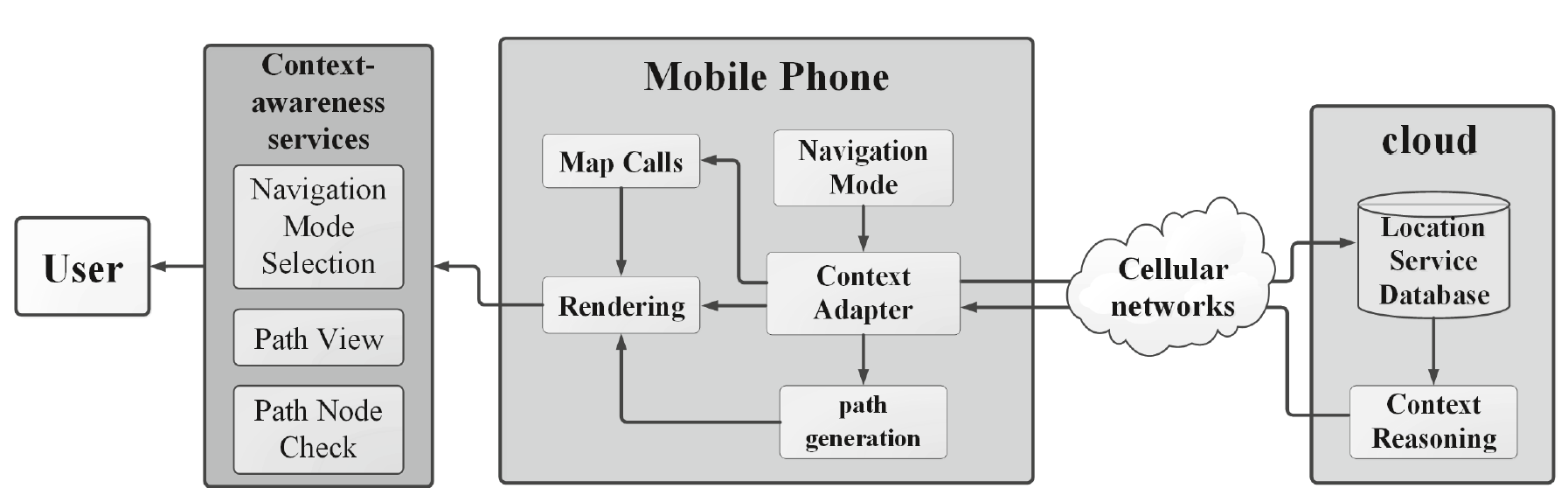}
    \caption{Overview of System Workflow}
\end{figure}

After the request is sent navigation, POI label is a scene in the first choice, then the path becomes a fool navigation scheme. By drawing a shortest path from the current location to a specified location on a floor plan, route guidance service is to provide users with a fool. To facilitate the users to access the destination location information by using different modes of transport, not only has many navigation modes, such as walking, by bus, to choose from, but also for each navigation mode to draw a path in this scenario. Users can query the node information, such as bus stations, subway stations, intersections, etc., through the path. First, find a user wants to know the building arrived at the scene pointed out that real estate. Then by interacting with three-dimensional model, further details of the building and deepening understanding. Architectural details of the building structure, type of area, sales prices and property information, is present in the two-dimensional image or text label. In order to bring a more intuitive browsing experience users can zoom or rotate a three-dimensional model by touch control during the interaction. Marbury provide context aware services such as navigation mode selection, path and path node checks. Figure 5. Identification and browsing in POI scene 4 modular structure and context-aware services.

\begin{figure}
    \includegraphics[width=18cm]{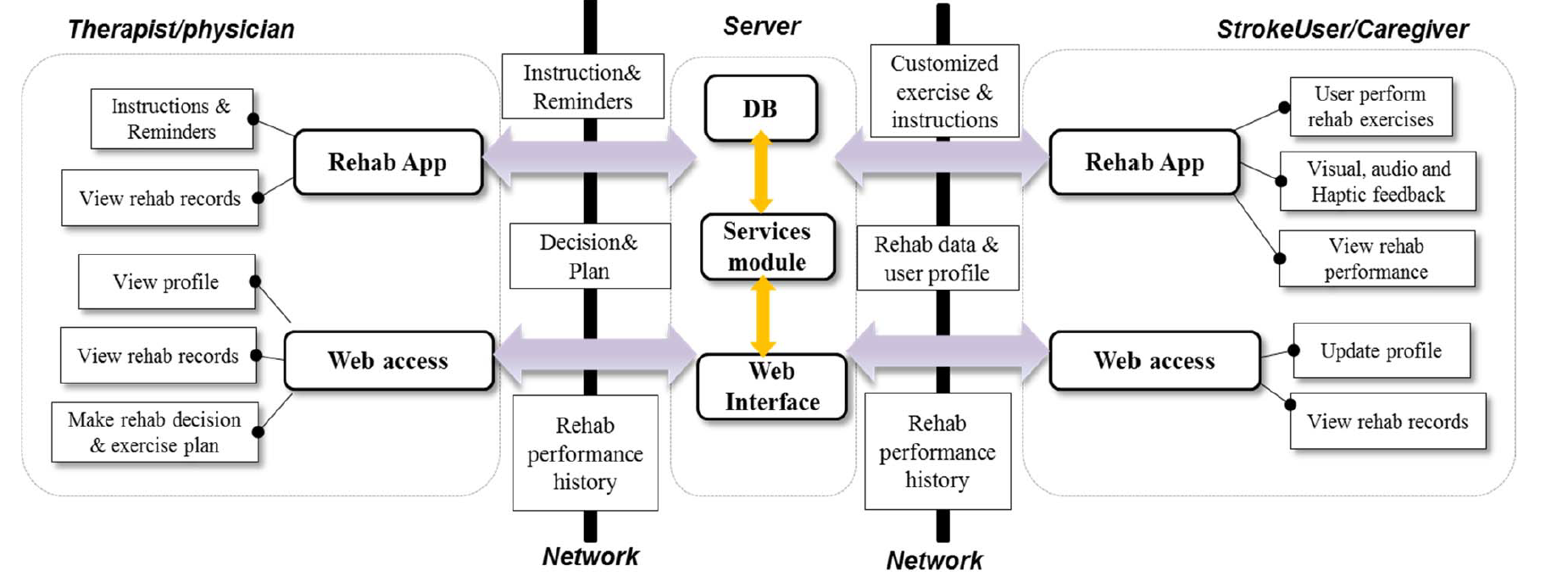}
    \caption{Face based Interaction System}
\end{figure}
This step is to produce the outline of the back and sides of the mask. Since the background is green, a hue threshold value is used to automatically distinguish between green area (body) from the green area (background). In order to obtain a clean side of the body contour, silhouette image in the back of the legs must be completely hidden behind the other front leg. In the image analysis, shown in FIG. 1, the back and sides of the contour image used. In order to facilitate the extraction of the body mask, participants wearing tight shorts and tank tops (F), and standing in front of a green background. In the following outline, arms participant leaves the body, feet shoulder width apart; and on the side, arms close to the body to the participants, with feet together. Focus and position of the camera is fixed. The generated body mask (e.g., the side and the back masks shown in Fig. 2) is further rotated to ensure that the body is in straight standing position. This is achieved by drawing a vertical line passing centroid computed using the smallest bounding rectangle containing the body mask) and continuously rotating the image until the maximum area symmetry is obtained on both sides of this vertical line (obtained when the difference in pixel count between left and right halves of the body mask separated by this vertical line is minimum).

\begin{figure}
    \includegraphics[width=12cm]{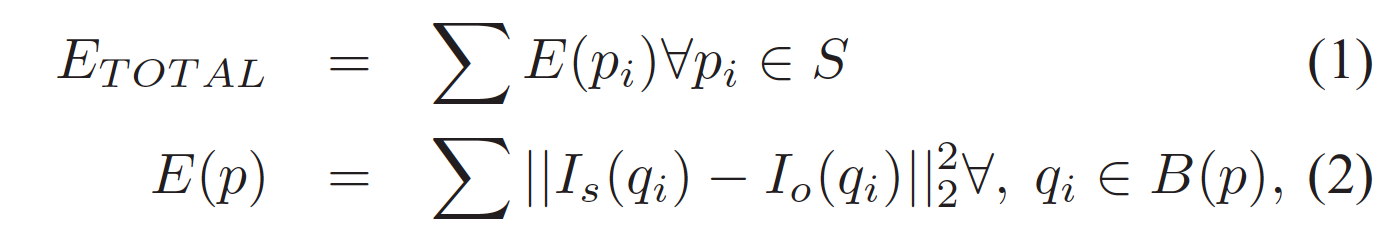}
\end{figure}

We support the real-time communication between the client and the server, the data will automatically store all the data streams in the database. Or, if you do not provide wireless network services, user data can be temporarily stored on the mobile device. On the network data stream may be different, the corresponding time sensitivity. Data, such as user profiles and exercise program, do not change very often, it does not affect the quality and quantity of stroke rehabilitation exercises. On the other hand, some of the data stream is a very sensitive time, including recovery record and intervention guidance, reminders and encouragement. These are very important to complete the design of the user stroke and exercise therapists have a better control of the patient's self-practice, so these data are needed in real-time delivery. In this system, the data stream is usually text-based, has a very small size, such as a few KB, so it is easy to ensure the real-time conversion, even in a wireless environment. The user can interact with the system in two ways, namely the proposed rehabilitation program, or computer-related Web access. Access methods may have different functions, depending on the user privileges assigned category. For example, by the mobile application, users can stroke rehabilitation exercises, view their practice and history, and access to visual, auditory and tactile feedback in real time during the exercise. However, the therapist can only view user profiles and resume recording, if they choose Web access method used to make decisions and rehabilitation exercise program.

\section{System Evaluation}
Recommendation mobile application has been tested in a variety of network conditions, to evaluate its performance and stability. Specifically, the four wireless connection, a 50MB / s Wi-Fi, 4G LTE and a very low speed wireless has been used in the experiment. The first three cases, the application will automatically send user data to the server in real time, and in the last case, the data will be stored in the local phone or tablet, and after flow whenever a wireless connection is available. The size of the data stream only a few KB, so it is clear that such a small data streams can be easily in real-time, even if conditions are not met in the radio. When a wireless connection is not good, the user data stored in the local reasons, as the fourth case, to ensure the security of the data stream. Loss of health-related data in the transmission process is unstable network we want to reduce the risk.

Changes have been developed two game programs and meet the different preferences, users from the wind. Specifically, the user can choose to play the game "monkey" game or "little astronaut" in the past a user can control a little monkey named "climb a tree" Looking for bananas and in the latter user control a small aerospace members perform spatial tasks. Figure 2 shows armstrokes two screenshots of the game. Game art is a fragment from a network resource [14,15]. Users through different rehabilitation exercise control of the game in animated characters. For example, the object of the forearm rotational movement is caused by the rotation of the monkey. In addition, real-time visual, audio and tactile feedback has been included in the game interactive and the user can select a different option in the game configuration. After each exercise test, the recovery results are displayed to the user, and automatically uploaded to the server. Users can also view their training history via the network. The data collected includes: date and time of exercise test; the test of time; the number of completed actions; motion distribution and performance. 

\begin{figure*}[t!]
    \centering
    \begin{subfigure}[b]{0.5\textwidth}
        \centering
        \includegraphics[height=1.2in]{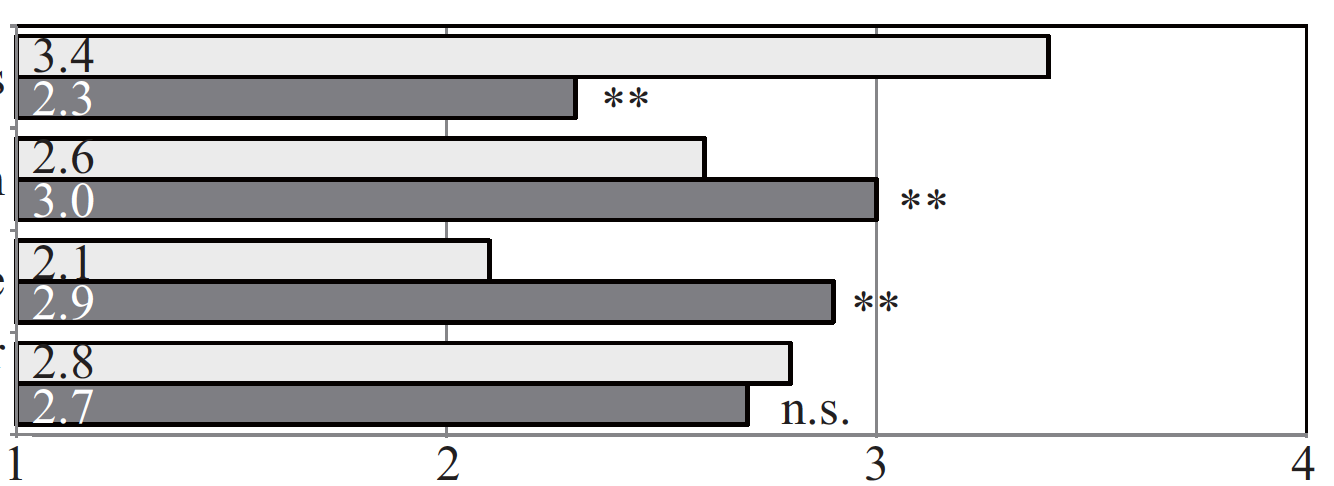}
        \caption{TSC System from \cite{11}}
    \end{subfigure}%
    ~ 
    \begin{subfigure}[b]{0.5\textwidth}
        \centering
        \includegraphics[height=1.2in]{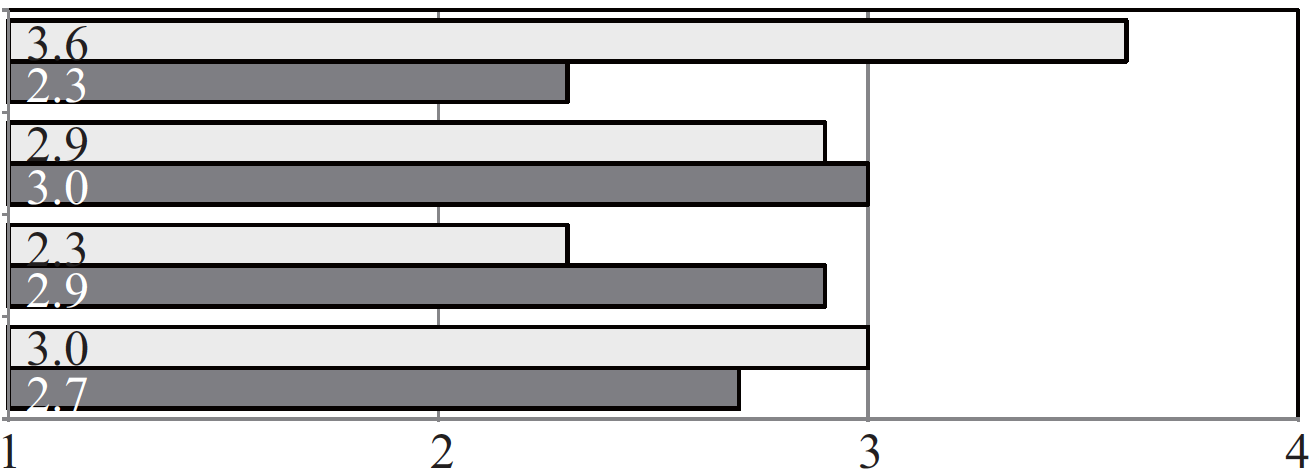}
        \caption{ISMO System from \cite{12}}
    \end{subfigure}
      
    \begin{subfigure}[b]{0.5\textwidth}
        \centering
        \includegraphics[height=1.2in]{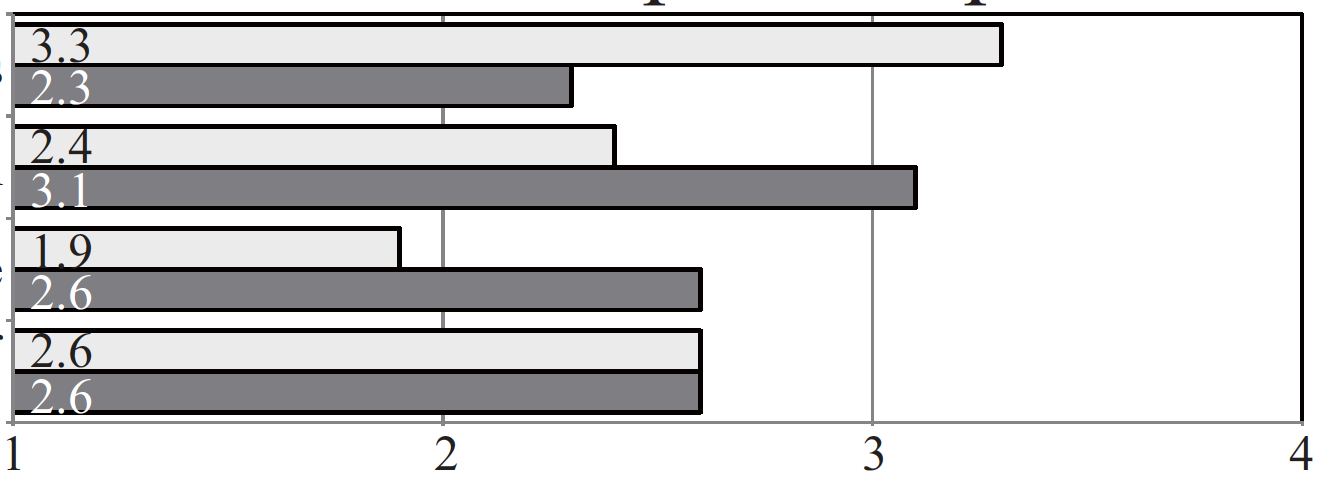}
        \caption{EUC System from \cite{03}}
    \end{subfigure}
        ~ 
    \begin{subfigure}[b]{0.5\textwidth}
        \centering
        \includegraphics[height=1.2in]{2.png}
        \caption{FCO System from \cite{08}}
    \end{subfigure}
    \caption{Quantified metrics by comparing with other VR based systems}
\end{figure*}

\section{Conclusion}
In this paper, we propose a new collaborative practice-based naming system and mobile devices to provide real-time support for the recovery. Through it, we can not only relates to stroke survivors, their caregivers, therapists and doctors play an important role in the rehabilitation process. As a feedback capacitor-based mobile application, called armstrokes, we proposed to provide rehabilitation through interactive games limb rehabilitation. Users can get real-time stroke of visual, audio and tactile feedback about their practice results, and see their performance. Caregivers, therapists and doctors can immediately monitor the rehabilitation process and practice outcomes for patients through the system, and then follow the patient through treatment such as reminders, encouragement and warning.

\bibliographystyle{asmems4}

\bibliography{asme2ejs}
\end{document}